\documentclass{epl}
\usepackage{epsfig}

\title{Pattern Formation in a Rotating Aqueous Suspension}

\author{
A. P. J. Breu \and
C. A. Kruelle \and
I. Rehberg
}

\institute{
Experimentalphysik V,
Universit\"at Bayreuth,
D-95440 Bayreuth, Germany
}


\begin{document}

\maketitle

\begin{abstract}
A novel pattern forming instability in a mixture of a granular material and
water in a horizontal rotating drum is experimentally investigated. The
particles accumulate in radial symmetric rings separated by pure water.
The transition between the homogeneous state and the structured state is
hysteretic.  The transition point is extrapolated from the growth rates of
the chain of rings.  The trajectory of a single particle is discussed to
estimate an upper boundary for the transition point.
\end{abstract}

\section{Introduction}

Granular media show an interesting and peculiar dynamic behavior
\cite{Jaeger:1996}, and are also of major importance in technological
processes. One of the standard problems of the powder technology is the
mixing of materials \cite{Miyanami:PowderTechHandbook}.  It is often
accompanied by pattern formation like stratification \cite{Makse:1997} or
segregation \cite{Frette:1997,Hill:1994,Dury:1999,Khakhar:1997}. The axial
and radial segregation in rotating-drum mixers with dry multicomponent
mixtures, like small and large particles or particles of different
densities, has been investigated. The different angles of repose of the two
species are considered as the segregation mechanism for the axial
segregation. The separation of large and small particles has recently been
observed in a mixture of particles dispersed in a lower density fluid
\cite{Jain:2001}. In the presence of a fluid the time for the pattern
formation is significantly decreased.

Granular media immersed in a fluid show a type of segregation even for
monodisperse particles \cite{Tirumkudulu:1999, Boote:1999,
Tirumkudulu:2000, Thomas:2001}.  The volume fraction of granular particles
in these experiments is much less than in the previous cases.  Furthermore,
the driving frequency of the drum is sufficiently high for the centrifugal
forces to become dominant.  The observed pattern is a chain of
circumferential rings of high particle concentration.  Tirumkudulu {\it et
al.} reported on this type of axial segregation in partially filled
horizontal cylinders \cite{Tirumkudulu:1999,Tirumkudulu:2000}. In both
cases they used suspensions of particles, which were neutrally buoyant.  If
the device was completely filled with the suspension, no instability was
observed at all.  The experiment was modeled  theoretically by Govindarajan
{\it et al.}\cite{Govindarajan:2001} including shear-induced diffusion of
particles, concentration-dependent viscosity and the existence of a free
surface. Boote and Thomas\cite{Boote:1999}, and Thomas {\it et al.}
\cite{Thomas:2001} reported similar experiments with particle densities
less and more dense than the surrounding fluid in a partially fluid-filled
cylinder.  For particles of a different density than the fluid, the free
surface is not necessary for the instability. Lipson observed this
phenomenon during crystal growth in a supersaturated solution
\cite{Lipson:2001}.

In this publication we present experiments on axial segregation of
monodisperse particles which have a higher density than the surrounding
fluid.  The experiments have been carried out in a drum, which was
completely filled with a sand-water mixture. We estimate an upper boundary
for the critical frequency for the onset of the pattern formation by the
numerical analysis of the trajectory of a single particle.

\section{Experiments}

\begin{figure}
\centerline{\epsfig{file=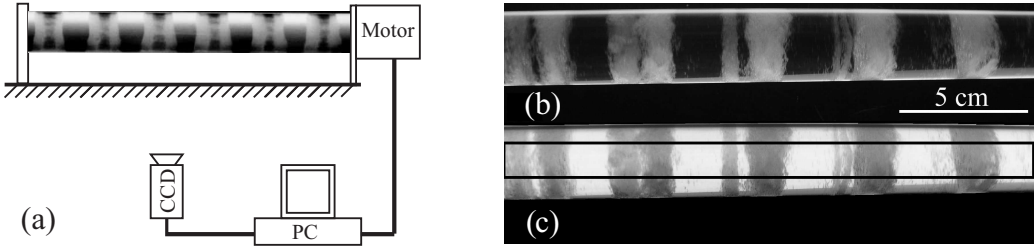}}
\caption{
(a) Experimental setup. (b) Photograph of the tube with the pattern.  A
top view of the tube is shown.  The sand is accumulated in the white
areas. (c) Corresponding transmitted light intensity, sand
accumulations appearing dark. The 5 pixel wide frame indicates the
area considered for further analysis.
}
\label{setup}
\end{figure}

The experimental setup is shown schematically in
\mbox{Fig.~\ref{setup}}(a).
%
%	Setup
%
The experiment consists of a horizontally aligned plexi\-glass tube, which
is $400\un{mm}$ long.  Its inner diameter measures $23.73\pm 0.05\un{mm}$.
Both ends are closed with plexi\-glass plugs. The tube is filled with \mbox
{2.5\%} (by volume) glass beads and \mbox{97.5 \%} water.  This is
sufficient to reach about \mbox{85\%} of the close packing for a monolayer
on a cylindrical surface.  The glass beads are ballotini impact beads with
a density of $2.45\un{g\,cm^{-3}}$ and diameters between $280\un{\mu m}$ to
$300\un{\mu m}$.

One end of the tube is attached rigidly to a dc-motor, which rotates the
tube around its cylinder axis.  The motor speed is controlled with an
accuracy of $10^{-4}$.  The control unit allows rotation frequencies
$f_\mathrm{rot}$ from $0\un{Hz}$ to $50\un{Hz}$ in steps of
$\frac{1}{60}\un{Hz}$ with a maximal acceleration of $1047\un{s^{-2}}$.
The axial sand distribution in the tube is measured via light transmission,
with a neon light placed behind the tube. A CCD camera, connected to a
frame grabber, detects the intensity of the transmitted light. Each frame
consists of a narrow strip (512 $\times$ 5 pixels) along the tube
displaying high sand concentrations as dark stripes in a bright surrounding
(Fig.~\ref{setup}(b) and (c)).

The experimental protocol consists of a stepwise increase or decrease of
the rotation frequency from $2.4\un{Hz}$ to $4.0\un{Hz}$
followed by a subsequent waiting period of $5\un{min}$, after which
transients have disappeared.
%
%
%	figure patterns
%
%
\begin{figure}
\centerline{\epsfig{file=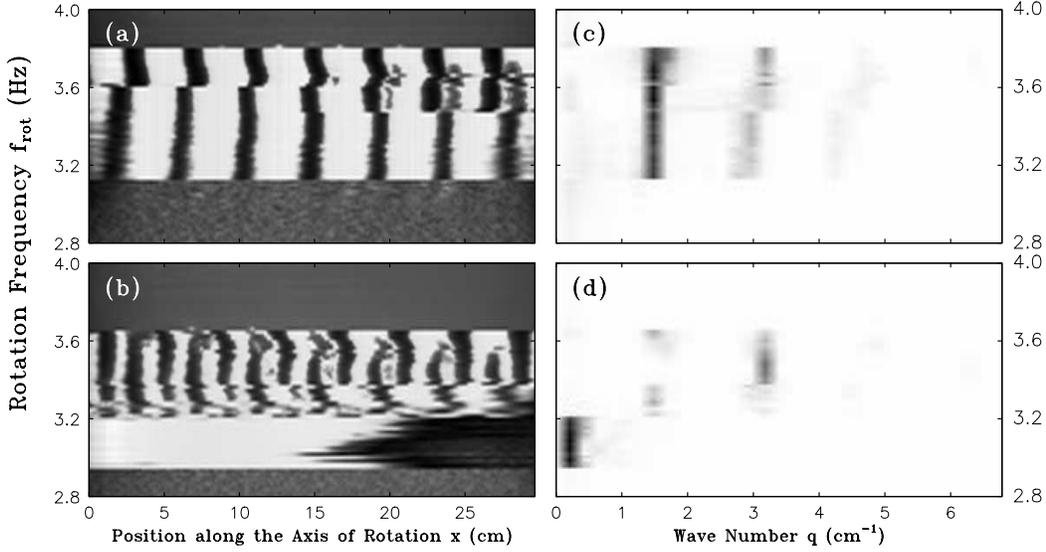}}
\caption{
Different patterns at quasistatic changes of the (a) increasing and (b)
decreasing rotation frequencies.  Each line corresponds to one measurement
of the sand distribution along the axes of rotation, which is an average of
20 images taken within $0.8\un{s}$.  These intensities are normalized by
the intensities obtained at zero rotation frequency, in order to prevent
errors due to inhomogeneous illumination. The lines are plotted with an
offset proportional to the rotation frequency to show the evolution of the
system. The frequency was swept between 2.4 Hz and 4.0 Hz. (c) and (d) show
the corresponding power spectra for increasing and decreasing rotation
frequencies.  Each line of (a) and (b) was transformed and plotted by the
same offset as before.
}
\label{patterns}
\end{figure}
No patterns are visible for high frequencies above $3.9\un{Hz}$ and
frequencies below $2.9\un{Hz}$ (\mbox{Fig.~\ref{patterns}}(a) and (b)).
At intermediate frequencies patterns appear, depending on the history of
the system.  Thus we display the intensities for increasing and decreasing
rotation frequencies in part (a) and (b) of \mbox{Fig.~\ref{patterns}}.

Alternatively we characterize the pattern by a discrete Fourier
decomposition of the lines (Fig.~\ref{patterns}). As soon as the rings
appear, well defined peaks show up in the power spectrum. In the case of
decreasing rotation frequencies the first peak is located at
$q_\mathrm{low}=1.49 \pm 0.05\un{cm^{-1}}$, which corresponds to a ring
spacing of $4.23 \pm 0.15\un{cm}$.  For frequencies below $3.5\un{Hz}$ the
first peak almost vanishes and a peak at
$q_\mathrm{high}=3.19 \pm 0.05\un{cm^{-1}}$ becomes dominant, due to a
splitting of the rings. The pattern returns to $q_\mathrm{low}$ for
frequencies between $3.4\un{Hz}$ and $3.25\un{Hz}$. Below $3.25\un{Hz}$ the
particles start to accumulate at the right hand side of the tube. Whether
the system chose the right or the left side could not be predicted in
advance.  For rotation frequencies below $2.9\un{Hz}$ the sand
distributed again homogeneously in the tube.  Increasing the rotation
frequency leads to a different scenario as shown in Fig.~\ref{patterns}(a)
and \ref{patterns}(c). There the characteristic ring spacing is almost
constant at $q_\mathrm{low}$.  No accumulation at a tube end was observed
even after waiting periods of 30 min.

\begin{figure}
\centerline{\epsfig{file=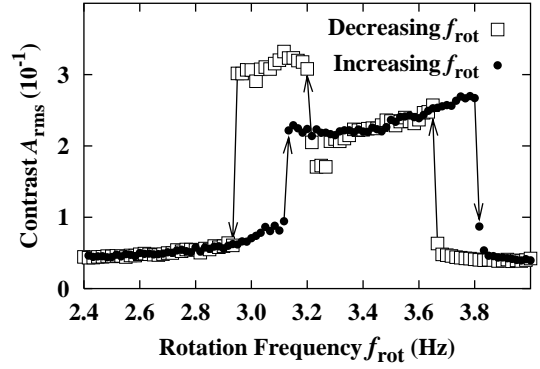}}
\caption{Amplitude of the pattern as a function of
the rotation frequency.}
\label{Amplitude}
\end{figure}

To get a more precise picture of the transition points, we take as an order
parameter the root mean square $A_\mathrm{rms}$ of the measured intensities
along the axes of rotation (Fig.~\ref{Amplitude}).
%
%
%	figure Amplitude
%
%
We observe two hysteresis loops, which are related to the transitions
between structured and unstructured states. These hysteresis loops indicate
that the transitions can be considered as subcritical bifurcations.

\begin{figure}
\centerline{\epsfig{file=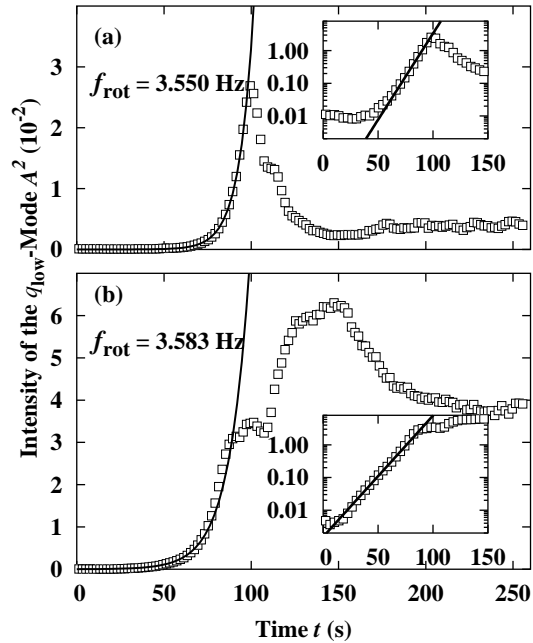}}
\caption{Growth of the amplitude of the pattern as
a function of time for two different rotation frequencies.  The solid
line represents the fit of an exponential function to the data.  The insets
show the regions of exponential growth on a semi-logarithmic scale.
}
\label{growth}
\end{figure}

In order to characterize the bifurcation around $3.7\un{Hz}$ further we now
concentrate on the temporal evolution of the instability. We assume
\begin{equation}
\partial_{t} A = \sigma A
\label{aeq}
\end{equation}
where $A$ describes the amplitude for a given wave number of the pattern,
and $\sigma$ the growth rate. To measure $\sigma$ as a function of the
driving frequency we start at a rotation frequency, where the granular
medium forms a homogeneous layer ($f_\mathrm{rot}=4\un{Hz}$). Then we
decelerate the motor suddenly to a frequency where the ring pattern is
stable ($f_\mathrm{rot} < f_\mathrm{rot,c}$).  Due to this sudden change,
the pattern will grow in time, as described by equation (\ref{aeq}) to
lowest order.  Figure \ref{growth} shows the growth of the amplitude in
time after a jump from $4.0\un{Hz}$ to $3.583\un{Hz}$
(Fig.~\ref{growth}(b)), and to $3.550\un{Hz}$ (Fig.~\ref{growth}(a)).  The
amplitude in this plot was obtained from the power spectrum by selecting
the mode $q_\mathrm{low} = 1.49\un{cm^{-1}}$, which is the first stable one
if the rotation frequency is decreased quasistatically (see
Fig.~\ref{patterns}(d)).  The solid line is the fit of an exponential
function to the data.  At higher amplitudes the growth is no longer
exponential. In fact, the amplitudes decays after $100\un{s}$
(Fig.~\ref{growth}(a)). This non-monotonic effect is caused by a
rearrangement of the pattern towards a higher wave number.
%
%	growth of the amplitude
%

\begin{figure}[h]
\centerline{\epsfig{file=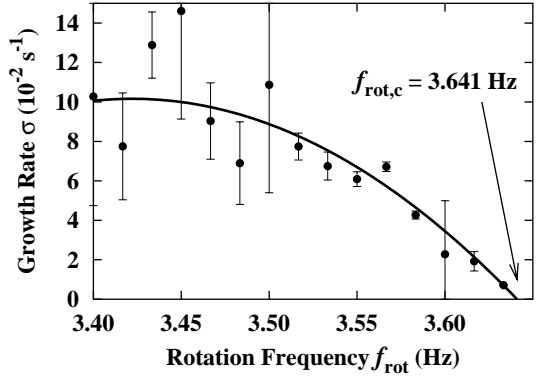}}
\caption{Growth rate as a function of driving frequency
$f_\mathrm{rot}$.}
\label{sigma}
\end{figure}

This quenching of the system was performed for final rotation
frequencies between 3.4 Hz and 3.9 Hz.  Above 3.65 Hz no growth of the
pattern was observed within a measurement time of 5 minutes. Below
$3.4\un{Hz}$ it was not possible to measure a growth rate for
$q_\mathrm{low}$. For frequencies between 3.633 Hz and 3.567 Hz the final
stable mode was $q_\mathrm{low}$. At lower frequencies this mode first grew
in time and decayed afterwards, being replaced by the higher stable mode
$q_\mathrm{high}$.  Nevertheless, it was also possible to extract a growth
rate for these frequencies, but with less accuracy, as indicated by the
error bars in Fig.~\ref{sigma}.

All measured growth rates are plotted as a function of the rotation
frequency in Fig.~\ref{sigma}.  For $f_\mathrm{rot}$ between 3.55 Hz and
3.65 Hz the growth rate of the mode $q_\mathrm{low}$ increases roughly
linearly with decreasing frequency, and saturates around 3.4 Hz, due to the
different stable mode $q_\mathrm{high}$, as seen in Fig.~\ref{growth}(a).
To model the increasing growth rate as well as the saturation, we fitted a
parabola to the data.  This parabola was then used to extrapolate the data
for $\sigma \to 0$. The corresponding critical rotation frequency is
$f_\mathrm{rot,c} = 3.641 \pm 0.035\, {\rm Hz}$.
%
%	growth rates
%
%
%	Discussion
%
\section{Discussion and conclusion}
In order to estimate an upper boundary for the critical rotation frequency
$f_\mathrm{rot,c}$ we analyze the motion of a single particle in the
laboratory frame, neglecting the interaction with the neighboring
particles. The calculation estimates the frequency for the detachment of a
particle from the wall, which is a necessary, but not a sufficient
criterion for pattern formation. A complete understanding of the
instability cannot be achieved by this simple calculation. At high rotation
frequencies the suspension is in rigid body rotation, and the particle
sticks to the inner wall of the tube.  Below a critical frequency the
particle detaches from the wall and falls on a parabola-like trajectory
described by the equation of motion
\begin{equation}
m \ddot{\vect{x}} = \vect{F}_\mathrm{G,res} +\vect{F}_\ab{S}
+\vect{F}_\mathrm{cb}\,.
\label{equation_of_motion}
\end{equation}
$\vect{F}_\mathrm{G,res}$ represents the force of gravity, reduced by
buoyancy:
\begin{equation}
\vect{F}_\mathrm{G,res} = \frac{4}{3} \pi r^{3}
\left(\rho_\ab{p}-\rho_\ab{f}\right) \vect{g}\,.
\label{gravitation}
\end{equation}
$r$ is the radius of the particle, $\rho_\ab{p}$ and $\rho_\ab{f}$ are the
density of the particle and the fluid, respectively.  The drag of the fluid
on the particle is modelled by a modified Stokes formula
\begin{equation}
\vect{F}_\ab{S} = - 6 \pi \eta r
\left(\vect{v}_\ab{p} -\vect{v}_\ab{f}\right) \cdot\lambda\,,
\label{stokes}
\end{equation}
where $\eta$ denotes the dynamic viscosity of the fluid.  $\vect{v}_\ab{p}$
and $\vect{v}_\ab{f}$ are the velocity of the particle and the fluid,
respectively, at the position of the particle.  For the fluid velocity
$\vect{v}_\ab{f}$ we assume that the fluid is in rigid body rotation.
Since the particle velocity is approximately equal to the fluid velocity,
the Reynolds number is close to zero.  $\lambda$ is the correction of the
Stokes formula for particles close to the wall, derived by Brenner for
vanishing Reynolds number \cite{Brenner:1961}:
\begin{equation}
\lambda = \frac{4}{3} \sinh \alpha \sum_{n=1}^{\infty}
\frac{n(n+1)}{(2n-1)(2n+3)}
\times\left[\frac{2 \sinh (2n+1)\alpha + (2n+1) \sinh
2\alpha}{4\sinh^2 (n+\frac{1}{2})\alpha -(2n+1)^2 \sinh^2
\alpha}-1\right]\,,
\label{wallcorrection}
\end{equation}
where the particle radius dependence which is taken care of by $\alpha$ is
defined as
\begin{equation}
\alpha = \cosh^{-1}\left(\frac{d+r}{r}\right)\,.
\end{equation}
$d$ is the distance of the sphere's surface to the wall.

The third force is caused by the radial pressure distribution of the fluid
in rigid body rotation:
\begin{equation}
\vect{F}_\mathrm{cb} = - \frac{4}{3} \pi r^{3} \rho_\ab{f}\, \omega ^{2}
\vect{x}\,,
\label{cbf}
\end{equation}
the centrifugal buoyancy \cite{Batchelor:Intro}.

While the detachment point can be calculated analytically \cite{Breu:PhD},
the path of the particle is obtained by numerical integration of the
equation of motion (\ref{equation_of_motion}).  Note that the particle, at
the detachment point, has a distance of $\tilde{R}=R-r-\delta$ from the
center of the tube.  $R$ is the inner radius and $\delta$ the surface
roughness of the tube. It is necessary to include the surface roughness,
since the wall correction (\ref{wallcorrection}) becomes singular for
vanishing distance $d$.  We have chosen $\delta=1\, \mu\mathrm{m}$, which
is a typical value for cast processes.

\begin{figure}
\centerline{\epsfig{file=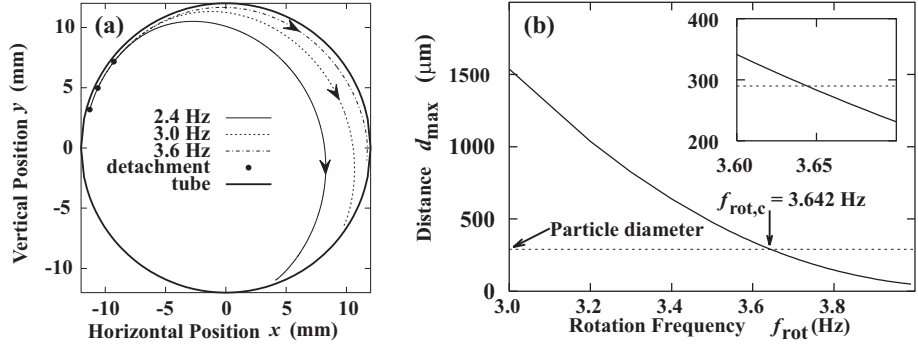}}
\caption{(a) Trajectories of a particle falling in a
viscous fluid for different rotation frequencies.  The fluid is
assumed to perform a rigid body rotation.
(b) Maximal distance to the tube wall of a
particle falling in a viscous fluid as a function of rotation
frequency.  The dashed line is drawn at the particle diameter.  The
inset is a close-up of the critical region.}
\label{simulation}
\end{figure}

Starting from the detachment point, we integrate the equation of motion
(\ref{equation_of_motion}) until the particles touched the wall of the tube
again. Three different trajectories are shown in Fig.~\ref{simulation}(a).
From these trajectories we calculate the distance of the particle from the
tube wall as a function of time.  Each curve has a unique maximum
$d_\mathrm{max}$, which depends on the rotation frequency of the tube.
We determined numerically this maximal distance for different rotation
frequencies (Fig.~\ref{simulation}(b)).  For the rearrangement of the
particles, which initially tend to form a close packed monolayer, into
separated rings, some individual particles have to jump over their
neighbors.  In order to do this they have to achieve at least a distance
$d_\mathrm{max}$ equal to one particle diameter.  If the distance is less
than the diameter, particles will not be able to pass their neighbors.
Therefore the critical frequency is found at the distance $d_\mathrm{max}$
equal to one particle diameter $2r$. By this we obtain a critical rotation
frequency of $3.642 \pm 0.060\un{Hz}$, which is in agreement with the
experimental value of $3.641 \pm 0.035\un{Hz}$.

A detailed theoretical investigation of the interaction between sand and
water leading to the ring pattern remains to be done.  A two-fluid approach
might be difficult, since the primary flow is not trivial
\cite{Lange:private}.  A molecular-dynamics simulation
\cite{Kalthoff:1997}, taking into account both the trajectories of the
grains and the fluid flow, seems to be more promising.

\acknowledgments
We like to thank S. Auma\^{\i}tre, V. Frette, and M.A. Scherer for many
interesting and inspiring discussions.  This work has been supported
by the Deutsche Forschungsgemeinschaft RE-588/12.

\bibliographystyle{unsrt}
%\bibliography{gm}

\end{document}